\documentclass[a4paper]{article}
\usepackage{amsmath}
\usepackage{graphicx}
\begin{document}
\title{The $0^{-+}$ and $0^{++}$  glueballs\\ and the multiplets
including them\footnote{Presented at the Workshop "Supersymmetries and Quantum Symmetry",
Dubna 2007}}

\author{Micha{\l} Majewski\thanks{e-mail: m.majewski@merlin.phys.uni.lodz.pl}\\
Department of Theoretical Physics II, University of Lodz} \maketitle
\begin{abstract} The properties of the $0^{++}$ and $0^{-+}$ meson
multiplets are discussed. Quoted are the $0^{++}$ and $0^{-+}$
glueball masses determined from data fit.

\end{abstract}

\section{Introduction}
The existence of glueballs is predicted by QCD, but to day none is
definitively established. The main problem is their identification.

There are predictions of lattice QCD simulation \cite{Lat} for the
lowest mass glueball of given signature $J^{PC}$. For the $0^{++}$
and $0^{-+}$ glueballs they are:
   $m_{0^{++}}\simeq1.5 GeV$ \quad and \quad $m_{0^{-+}}=2.3\div2.5GeV$.

There are also experimental statements. At present, the $f_0(1500)$
and $\eta(1405)$ mesons are allowed as glueball dominated mixtures
with isoscalar $q\bar{q}$ states \cite{PDG}. The data confirm
lattice prediction for the $0^{++}$, but they drastically disagree
with the prediction for the $0^{-+}$ glueball. This makes a trouble
as people trust the lattice predictions.

Of course, we may look for another field theoretical approach to
describe the data. Such description already exists. Faddeev, Niemi
and Widner proposed recently a topological model of the glueball as
closed flux tube \cite{Fad}. The model predicts degeneracy of the
$0^{++}$ and $0^{-+}$ glueball masses and admits the region
$1.3\div1.5$GeV, where they are really observed.

However, there is also a problem of exploiting data. Both the
$f_0(1500)$ and $\eta(1405)$ have been chosen as glueball candidates
from among a few isoscalar mesons on a basis of qualitative
information about their production and decay (e.g., "gluon rich
environment", or "flavor independence"). Although the predictive and
verification power of such procedure is not high, it is the only
generally accepted method of the glueball identification. However,
such a procedure exploits only part of the accessible data which can
be used.

Overpopulation of a nonet is an important signal of the glueball.
The glueball should mix with the isoscalar nonet states. All
properties of the decuplet which arises this way, including mixing
between three isoscalar states, can be described entirely by the
masses. Hence, the data on the masses can be used to determine
parameters of the glueball.

\newpage

\section{Exotic commutators and master equations}
We assume that the following set of exotic commutators vanishes
\cite{MT}:
\begin{equation*}
\left[T_\alpha,\frac{d^jT_\beta}{dt^j}\right]=0,\quad
(j=1,2,3,...) \label{0.1}
\end{equation*}
where $T$s are $SU(3)_F$ generators, $t$ is time and
$(\alpha,\beta)$ is an exotic combination of indices; that means
that $T_\alpha$, $T_\beta$ are chosen such that operator
$[T_\alpha,T_\beta]$ does not belong to the octet representation.
These equations are basic for the model of exotic commutators (ECM).
They can be transformed \cite{enc} into the system of
\textbf{master equations (ME)}:\\
\begin{equation*}
\langle{z_8}\mid{\hat{(m_c^2)}^j}\mid{z_8}\rangle = \frac{1}{3}a_c^j
+ \frac{2}{3}b_c^j, \quad (j=1,2,3,...) \label{0.2}
\end{equation*}\\
where $\hat{m_c^2}=\hat{m^2}-i\hat{m}\hat{\Gamma}$ is complex-mass
squared operator and $\hat{m}$ and $\hat{\Gamma}$ are hermitean and
commute. The operator $\hat{m_c^2}$ can be diagonalized and has
orthogonal eigenfunctions. $|z_8\rangle$ is the isoscalar octet
state, $a_c$ is the isovector particle mass squared, $ b_c=2K_c-a_c$
and $K_c$ is the mass squared of the isospinor particle.

The substitution $|z_8\rangle=\Sigma
 l_i|z_i\rangle$ ($\Sigma |l_i|^2=1$), where $|z_i\rangle$ are
 isoscalar physical states (i=1,2 -- for the nonet and i=1,2,3 --
 for the decuplet) transforms ME into a system of linear equations
 with respect to octet contents $|l_i|^2$:
\begin{equation*}
    \Sigma |l_i|^2z_i^j=\frac{1}{3}a_c^j+\frac{2}{3}b_c^j, \quad (j=0,1,2,3,..)
    \label{2.10}
\end{equation*}
where the equation for $j=0$ takes into account the normalization of
$l_i$s. To find $|l_i|^2$ we need at least two equations -- for the
nonet and three -- for the decuplet. Any additional equation must
comply with the solution and thus sets requirement on the masses. We
can choose the number of equations for the nonet and decuplet such
as to have just one complex mass formula. The $|l_i|^2$s being the
solution of the ME are expressed by the complex masses, but they
must be: $1^0$ real, $2^0$ positive.

$1^0$. The condition imposes a linear dependence between the widths
and masses (straight \textbf{flavor stitch line (FSL)}) -- the same
for the nonet and its decuplet extension. The dependence reduces the
solution $|l_i|^2$ (not quoted here) and the complex mass formulae
to the form of the real mass meson multiplets:\\
$(x_1-a)(x_2-a)+2(x_1-b)(x_2-b)=0$ -- for the nonet
\footnote{Note
that there are three kinds of the nonets:
  Gell-Mann -- Okubo (GMO), Schwinger (S) and Ideal Mixing (I) \cite{Sum}.
 The ME system giving one mass formula points out the S nonet.} and\\
$(x_1-a)(x_2-a)(x_3-a)+
    2(x_1-b)(x_2-b)(x_3-b)=0$ -- for the decuplet.\\
Here $x_i$ are the masses squared of the isoscalar particles $z_i$.

$2^0$. This requirement, together with the mass formula, defines the
\textbf{mass ordering rule (MOR)} as another
condition for the existence of the multiplet:\\
$x_1<a<x_2<b$ \quad or $ \quad a<x_1<b<x_2 $ -- for the nonet and\\
$x_1<a<x_2<b<x_3$ -- for the decuplet.

Rectilinearity of the FSL follows from the flavor symmetry, but the
slope $k_s$ of  the line is not determined by ME -- it can be found
only from data fit. The determination is not always possible, but in
all cases where it can be done the slope is almost the same:
$k_s\simeq-0.5$.

However, if the mass of a particle is smaller of about 1.5GeV its
(m,$\Gamma$) coordinates may not comply with FSL, because the decay
may be suppressed by some additional "kinematical" (non-flavor)
mechanism \cite{Sum}. The point with such coordinates lies below the
FSL. The solutions of ME include also relations between the
imaginary parts of the complex masses. They have the form of the
mass formulae but are satisfied by data worse than FSL.

The definition of the multiplet is based entirely on the mass
formula, hence it is independent of the "kinematical" breaking of
the widths. Also the mixing matrix of the decuplet isoscalar states
does not depend on the widths; it is determined by the masses via
solution $|l_i|^2$ and is real, even for complex masses.

\section{Spectra of multiplets}
The independence of the definition of the multiplet on the widths
proves very important for the question of nature of the $0^{++}$
mesons in the 1GeV region. The width argument against their
$q\bar{q}$ structure is not valid and there is no need for
introducing the four quark states for them \cite{enc}. Strong
suppression of the $f_0(980)$ decay is just the manifestation of
some ("kinematical") suppression mechanism. Hence, the scalar mesons
from the 1GeV region belong to the common $q\bar{q}$ nonet, but the
energy dependence of the phases $\delta_{J=0}^{I=0}$ and
$\delta_{J=0}^{I=1}$ does not reflect the properties of the flavor
interaction.

The $0^{++}$ nonet includes \cite{enc} ($f_0(1710)$ is pointed out by the S mass formula):\\
$a_0(980)$,\quad $K_0(1460)$,\quad $f_0(980)$,\quad $f_0(1710)$;
\quad (the MOR is:\quad $x_1<a<x_2<b$)
Other $0^{++}$ mesons (not included into the nonet) constitute the decuplet:\\
$a_0(1460)$,\quad $K_0(1950)$,\quad $f_0(1370)$,\quad
$f_0(1500)$,\quad $f_0(2200)/f_0(2330)$,\\
where  $f_0(1500)$ is the glueball candidate \cite{Ams1}. The mass
domains of these multiplets overlap, but there is no mixing between
their states.

There exists one more meson observed below the mass scales of these
two multiplets -- the $\sigma(600)$ one. This meson cannot join the
nonet to form a decuplet (compare MORs of the nonet and decuplet).

Also the $0^{-+}$ mesons form a nonet and decuplet. Observe, that
the old nonet\\
  $\pi$,\quad $K$,\quad $\eta$, \quad  $\eta'$ \quad is the only known GMO one.
The decuplet comprises \cite{GMM}:\\
     \quad $\pi(1300)$,$\quad K(1450)$,\quad $\eta(1295)$, $\quad\eta(1405)$,
    $\quad\eta(1475)$,\\
where $\eta(1405)$ is recognized as glueball candidate \cite{PDG},
\cite{Ams2}.

Hence, both the $0^{++}$ and the $0^{-+}$ mesons form the same
sequences of multiplets. Some of the masses are not exactly known,
but this does not disturb the general picture. Notice, that not only
the sequences of the multiplets are similar -
also the inner structures of the decuplets are: \\
-- the physical mesons $f_0(1500)$ and $\eta(1405)$ which are
dominated by glueball states are settled down just between the
remaining
isoscalar mesons dominated by $N$ and $S$ quark states,\\
-- both decuplets are built up of the excited $q\bar{q}$ states,
hence both glueballs mix with the excited ${(q\bar{q})_{isoscalar}}$
states.

The latter property suggests affinity of the glueball with the
excited states. This is suggested especially by the mixing of the
$0^{++}$ glueball; its mass belongs to the region where the nonet
ground states and the decuplet excited states are overlapping, but
the glueball prefers mixing just with the excited $q\bar{q}$ states.

The lack of data does not allow us to extend this comparison to
higher masses.\\

The $0^{-+}$ and $0^{++}$ mesons form the parity related multiplets
(nonets and decuplets). The sequences of these multiplets differ
only due to existence of the scalar meson $\sigma(600)$ which has no
pseudoscalar counterpart. But the nature of this meson is a matter
of discussion. Several authors suggest that its nature is different
from the nature of other mesons. By abandoning the $\sigma(600)$ we
find that $0^{-+}$ and $0^{++}$ mesons form parity related
\textbf{spectra of flavor multiplets}.

\section{Glueballs}
The masses of the glueballs are:\quad $m_{G^{-+}}=1.369 GeV$ and
\quad $m_{G^{++}}=1.497 GeV.$\\
Approximately $m_{G^{++}}-m_{G^{-+}}=m_\pi$; in any case \cite{GMM}
they satisfy
$ m_{G^{-+}}<m_{G^{++}}.$ \\
Clearly, the value of \quad $m_{G^{-+}}$ \quad supports the
prediction of the closed flux-tube model \cite{Fad} and is far from
the lattice QCD prediction \cite{Lat}.

\section{Conclusions}
1. The mesons $f_0(1500)$ and $\eta(1405)$ can be understood as
glueball dominated. \\
2. The nonet and decuplet states with the same $J^{PC}$ do not mix.\\
3. The glueballs have affinity to the excited $q\bar{q}$ states.\\
4. The mesons $0^{++}$ and $0^{-+}$ form parallel (parity related)
spectra of multiplets.\\
5. Perhaps the $\sigma(600)$ meson does not belong to the flavor
spectrum of $0^{++}$ multiplets.

\textit{Acknowledgments} I would like to thank Prof. E. Ivanov and
all organizers of QSQ'07 Workshop for invitation and excellent work.
I thank Profs S. B. Gerasimov and V. A. Meshcheryakov for valuable
discussions. This work was supported by the B--I Fond and the
University of Lodz (grants 690 and 795).

\end{document}